\documentstyle[aps,twocolumn,epsbox]{revtex}
\begin{document}
\draft
\title{
  A class of Heisenberg models with the orthogonal dimer    
  ground states 
}
\author{
  Kazuo Ueda and Shin Miyahara
}

\address{
  Institute for Solid State Physics, University of Tokyo,\\
  7-22-1 Roppongi, Minato-ku, Tokyo 106-8666, Japan
}

\date{\today}

\maketitle

\begin{abstract}

  Extensions of the Shastry-Sutherland model are possible 
  in various ways. In particular, it is possible to construct
  a natural model in three dimensions which has the exact 
  dimer ground state.  Recently found spin gap system 
  SrCu$_2$(BO$_3$)$_2$ has this structure. 
  The exchange constants between the layers is expected to
  be smaller than the intra-layer couplings. However,
  the exactness of the dimer state for the three dimensional
  structure is important to understand why magnetic properties of
  SrCu$_2$(BO$_3$)$_2$ are described well by the two dimensional
  model.

\end{abstract}

\pacs{75.10.-b, 75.10.Jm, 75.30.Kz}


\narrowtext


Recently Kageyama et al found a new two dimensional spin gap system 
SrCu$_2$(BO$_3$)$_2$\cite{kageyama}.  
The crystal structure of SrCu$_2$(BO$_3$)$_2$
is tetragonal and all Cu$^{2+}$ ions with a localized spin $S=1/2$ are
located at crystallographically equivalent sites. The two dimensional 
layers containing the Cu$^{2+}$ ions are separated by planes of
Sr$^{2+}$ ions. The two dimensional Heisenberg model with the
nearest neighbor, $J$, and the next nearest neighbor couplings, $J^{'}$
is topologically equivalent to the Shastry-Sutherland
model which was found nearly twenty years ago\cite{shastry}.  
  
The model shows a quantum phase transition from the
dimer ground state to the N\'{e}el ordered state\cite{miyahara}.  
The critical value for the transition is determined to be 
$(J^{'}/J)_c=0.7 \pm 0.01$.  From the magnitude of the spin gap
and the Weiss constant the coupling constants for SrCu$_2$(BO$_3$)$_2$
are determined as $J= 100$ K and $J^{'} = 68$ K.  Thus the ratio
between the two coupling constants is very close to the
critical value. It is shown that the closeness to
the transition point is the origin of the unusual temperature 
dependence of the susceptibility. 

Another novel feature of SrCu$_2$(BO$_3$)$_2$ is the magnetization
plateaus at one quarter and one eighth of the full Cu$^{2+}$ moment.
Occurrence of the plateaus originates from the nearly localized 
nature of the triplet excitations in the peculiar lattice 
structure of the model\cite{miyahara}.
Crystallization of the excited triplets at certain magnetization
is the origin of the magnetization plateaus.   

In this letter we will show that generalization of the 
Shastry-Sutherland model is possible in several ways. 
In particular, we can construct several models in three 
dimensions which have the exact dimer ground states. 
Interestingly enough, the three dimensional structure of 
SrCu$_2$(BO$_3$)$_2$ has a structure in this class. Although
three dimensionality of this compound may not be so 
important because of small interlayer couplings,  
it is certainly a helpful fact to understand why the analyses 
by the two dimensional orthogonal dimer model work so 
well for this compound.  

We start our discussion from two examples in one dimension. 
The one dimensional version of the Shastry-Sutherland model is
shown in Fig.1(a).  The bonds denoted by the thick solid lines 
define a unique covering of the spins and the dashed lines 
are the bonds connecting dimers. We use $J$ for the
coupling constant in the dimers and $J^{'}$ for the inter
dimer couplings. 

The dimer state is defined by 
\begin{equation}
   |\Psi\rangle = \prod_a | s\rangle_a
\end{equation}
where $| s\rangle_a$ is the singlet state for the dimer bond
specified by index $a$ which runs over all dimer bonds.
For this model the dimer state is an exact
eigenstate for any $J^{'}/J$ and the ground state
for small $J^{'}/J$\cite{ivanov}.  
To prove this, let us consider $J^{'}$ bonds 
connecting two dimers, Fig.2(a).
It is easy to see that all matrix elements vanish for the dimer 
state: 
\begin{equation}
   J^{'} ({\bf s}_1+{\bf s}_2) \cdot {\bf s}_3 
     | s\rangle_a | s\rangle_b = 0\ .  
\end{equation}
The result holds independent of the magnitude of the spin, $S$.

When we notice that the matrix elements vanish because 
of the difference in parities between the singlet and the 
triplet, it is easy to see that the  
same is true for any dimer pairs when they are 
orthogonal.  For the model of Fig.1(a), the neighboring dimers are 
orthogonal but in the same plane.  A different one dimensional 
Heisenberg model with the exact dimer ground state may be constructed 
by making one type of dimers out of the plane, Fig.1(b).  In this 
paper we define the bonds between the orthogonal dimers of this type  
as $J^{''}$ bonds, Fig.2(b). It is straightforward to confirm that
all matrix elements of the $J^{''}$ bonds vanish:
\begin{equation}
   J^{''} ({\bf s}_1+{\bf s}_2) \cdot ({\bf s}_3+ {\bf s}_4)
     | s\rangle_a | s\rangle_b = 0\ . 
\end{equation}
The present model is topologically equivalent to the spin ladders
with the diagonal couplings of the same amplitude
discussed by Gelfand\cite{gelfand}.

It is obvious that any combination of the two types of orthogonal 
dimers may have the dimer state as the exact ground state.  
In particular, the dimer state is the exact ground state 
even for a random distributions of the two dimer configurations
when the ratios $J^{'}/J$ and $J^{''}/J$ are sufficiently small.
Furthermore, random mixture of spins with different $S$ is possible
so long as the spins of each dimer have the same $S$. Distance between
the dimers can be different as long as neighboring dimers are 
orthogonal.

The Heisenberg model for each layer of SrCu$_2$(BO$_3$)$_2$
is shown in Fig.3.  This model can be considered as a
decorated square lattice with orthogonal dimer bonds in 
the two sublattices.  A different model with the exact 
dimer ground state is obtained by using dimer bonds 
out of plane in one sublattice.  Any combination of dimers
along three orthogonal directions is possible so long as the
neighboring dimers are orthogonal.

Another type of extension is possible even with only orthogonal
dimers in the plane.  Let us consider regularly depleted square 
lattice.  For Calcium Vanadates, a series of compounds 
CaV$_n$O$_{2n+1}$ with $n=2, 3, 4$ have been synthesized.  
These Vanadates have layered
structures and in each layer the spins of V$^{4+}$ ions form
the structure of the 1/(n+1)-th 
regularly depleted square lattice. By decorating the depleted 
square lattice we obtain a set of Heisenberg models with 
the exact dimer ground states for any $S$.  Two examples of
this class are shown in Fig.4. 

Now we proceed to three dimensions. The real three 
dimensional structure of SrCu$_2$(BO$_3$)$_2$ consists of
CuBO$_3$-layers and Sr-layers.  The CuBO$_3$-layers stack
alternately as is shown in Fig.5 \cite{structure}.  
It is clear that the spin-$S$ Heisenberg model of this 
structure has the exact dimer ground state for small 
$J^{'}/J$ and $J^{''}/J$.
Again we can construct a broad class of three dimensional
Heisenberg models by considering the triad of the
orthogonal dimers.

Strictly speaking, for SrCu$_2$(BO$_3$)$_2$
the dimers in each plane are not in 
the same plane: the plane of horizontal dimers is 
slightly shifted from the plane of vertical dimers. 
Because of this shift, the distance between Cu$^{2+}$ 
ions of the adjacent dimers along the $c$-axis is 3.5930 \AA\,  or
4.2325 \AA. On the other hand, the nearest-neighbor distance and
the next-nearest-neighbor distance in each plane
are 2.9046 \AA\, and 5.1316 \AA.
Actually the Cu$^{2+}$-Cu$^{2+}$ distances between the planes 
are shorter than the next-nearest-neighbor distance in the plane.   
In spite of this fact, we expect that $J^{'}$ is much
bigger than  $J^{''}$ since the dominant path of the 
superexchange is through the molecular orbital of BO$_3$.
On the other hand, Sr$^{2+}$ ion has a closed shell.
Therefore the two dimensional orthogonal dimer model
is a good starting point for SrCu$_2$(BO$_3$)$_2$.
However the exactness of the dimer ground state for the
three dimensional model is the reason why magnetic properties
of SrCu$_2$(BO$_3$)$_2$ are described well by
the two dimensional model.

It is our great pleasure to thank Dr. H. Kageyama for many 
helpful discussions.  We have also enjoyed illuminating discussions
with Dr. B.S. Shastry.

\begin{figure}
  \begin{center}  
    \psbox[width=6cm]{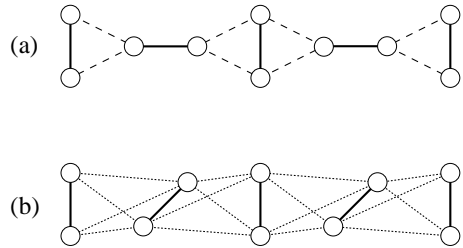}
  \end{center}
  \caption{
    Two examples of one dimensional orthogonal dimer
    models.
    }
  \label{fig:1D}
\end{figure}

\begin{figure}
  \begin{center}  
    \psbox[width=6cm]{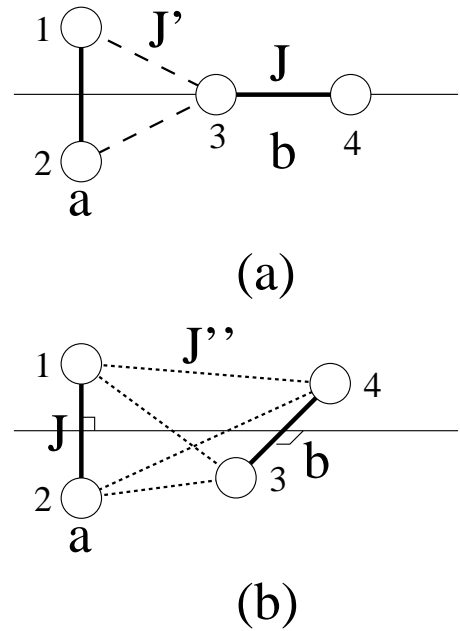}
  \end{center}
  \caption{
    Two configurations of orthogonal dimers.
    }
  \label{fig:units}
\end{figure}

\begin{figure}
  \begin{center}  
    \psbox[width=6cm]{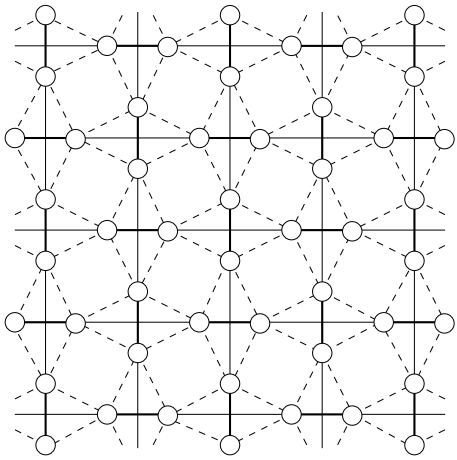}
  \end{center}
  \caption{
    The model for SrCu$_2$(BO$_3$)$_2$ : Two dimensional
    orthogonal dimer model.
    }
  \label{fig:2D}
\end{figure}

\begin{figure}
  \begin{center}  
    \psbox[width=6cm]{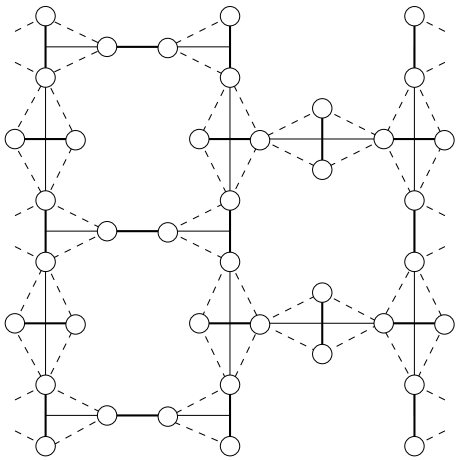}

    \vspace*{1cm}
    
    \psbox[width=6cm]{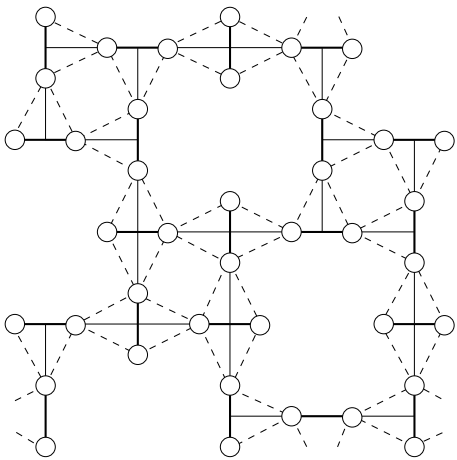}
  \end{center}
  \caption{
    Two examples of the regularly depleted square lattices with
    decoration of orthogonal dimers.
    }
\label{fig:2Ddpltd}
\end{figure}

\begin{figure}
  \begin{center}  
    \psbox[width=6cm]{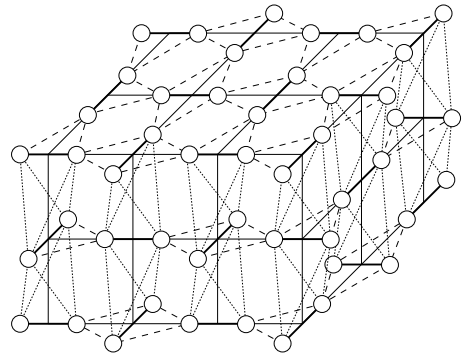}
  \end{center}
  \caption{
    Three dimensional orthogonal dimer model.
    }
  \label{fig:3D}
\end{figure}


\begin{thebibliography}{99}

\bibitem{kageyama} H. Kageyama, K. Yoshimura, R. Stern,
  N.V. Mushnikov, K. Onizuka, M.Kato, K. Kosuge, C.P. Slichter,
  T. Goto, and Y. Ueda: Phys. Rev. Lett. (1999) in press.
\bibitem{shastry} B.S. Shastry and B. Sutherland: Physica B
{\bf 108} (1981) 1069.
\bibitem{miyahara} S. Miyahara and K. Ueda:
Phys. Rev. Lett. (1999) in press.
\bibitem{ivanov} N.B. Ivanov and J. Richter: 
Phys. Lett. A {\bf 232} (1997) 308.
\bibitem{gelfand} M.P. Gelfand: Phys. Rev. B {\bf 43} (1991)
8644.
\bibitem{structure} H. Kageyama: private communication.

\end{thebibliography}
\end{document}